\documentclass[twocolumn,showpacs,pre,aps,superscriptaddress]{revtex4}
\usepackage[latin1]{inputenc}
\usepackage{bm}
\usepackage{multirow}
\usepackage{amssymb}
\usepackage{amsbsy}
\usepackage{amsmath}
\usepackage{stmaryrd}
\usepackage{graphicx}
\usepackage{subfigure}
\usepackage{epsfig}
\usepackage{hyperref}
\makeatletter
\usepackage{pifont}
\makeatother

\begin{document}

\title{Boltzmann-type approach to transport in weakly interacting one-dimensional fermionic systems}

\author{Christian Bartsch}
\email{cbartsch@uos.de}
\affiliation{Fachbereich Physik, Universit\"at Osnabr\"uck,
             Barbarastrasse 7, D-49069 Osnabr\"uck, Germany}

\author{Jochen Gemmer}
\email{jgemmer@uos.de}
\affiliation{Fachbereich Physik, Universit\"at Osnabr\"uck,
             Barbarastrasse 7, D-49069 Osnabr\"uck, Germany}

\date{\today}

\begin{abstract}
We investigate transport properties of one-dimensional fermionic tight binding models featuring nearest and next-nearest neighbor hopping, where the fermions are additionally subject to a weak short range mutual interaction. To this end we employ a pertinent approach which allows for a mapping of the underlying Schr\"odinger dynamics onto an adequate linear quantum Boltzmann equation. This approach is based on a suitable projection operator method. From this Boltzmann equation we are able to numerically obtain diffusion coefficients in the case of non-vanishing next-nearest neighbor hopping, i.e., the non-integrable case, whereas the diffusion coefficient diverges without next-nearest neighbor hopping. For the latter case we analytically investigate the decay behavior of the current with the result that arbitrarily small parts of the current relax arbitrarily slowly which suggests anomalous diffusive transport behavior within the scope of our approach.
\end{abstract}


\maketitle

\section{Introduction} \label{sec-introduction}

Considerable effort has been dedicated to the investigation of the
transport behavior of one-dimensional quantum wires. While it is
accepted that the transport behavior of the electrons will be normal
(diffusive) if the electrons are coupled to the phonons of the
supporting lattice \cite{jaeckle1978}, there is a discussion on whether or not
electron-electron interaction alone will render the transport
diffusive, i.e., non-ballistic. If the electrons are subject to a
periodic potential, as implied by any standard tight binding model the
total electron momentum is not necessarily conserved and one would expect diffusive
transport. However, 1-d spinless fermionic models with nearest neighbor hopping and
mutual interactions are accessible by a Bethe ansatz, which means that very many
local conserved quantities exist, thus they are called "integrable" \cite{giamarchi2004}.
As follows from the Mazur inequality the transport will be ballistic if
there is a significant overlap of the current with any conserved
quantity thus leading to a finite Drude weight \cite{zotos1997,mazur1969}. Such a finite Drude
weight has been found for zero temperature and small interactions
(gapless regime) for all fillings \cite{shastry1990}. A finite Drude weight has also
been found for finite temperatures and all fillings but half filling \cite{zotos1997,mazur1969}.
Recently there have been papers arguing in both directions for half
filling and finite temperature (in the gapless regime): in favor of
diffusive transport (in spite of integrability)\cite{sirker2009,steinigeweg2010,grossjohann2010} and in favor of
ballistic transport \cite{zotos1999,benz2005,narozhny1998,heidrich2003,jung2007,alvarez2002,heidarian2007,prosen2011}. 

Furthermore, the question how transport behavior changes if integrability
is continuously broken has been addressed. In the low temperature regime
an argument based on bosonization shows that at least two non-commuting
umklapp processes are required for non-ballistic transport \cite{rosch2000}. In all
temperature regimes diffusion coefficients may scale as $\lambda^{-4}$
(as opposed to  $\lambda^{-2}$) with $\lambda$ quantifying the
integrability breaking term if the perturbation allows for the
construction of new approximately conserved quantities \cite{jung2006}.

In the paper at hand we investigate a 1-d model of spinless fermions
with nearest neighbor hopping (nn), next-nearest neighbor hopping (nnn)
and a short ranged, perturbatively weak particle-particle interaction.
In this model the nnn-term is the one that breaks integrability. We
address transport behavior by (approximately) mapping the quantum
dynamics onto a linear Boltzmann equation via a projection operator
technique \cite{breuer2007}. For simplicity all calculations are done at infinite
temperature. Within this framework we find concrete transport
coefficients and evidence that they diverge as the nnn-term goes to
zero. This suggests non-diffusive transport in the gapless regime at all
temperatures. The time scales, however, at which this "high-mobility"
behavior emerges become infinitely long. Results from (approximate)
projection techniques may be wrong if the projected subspace does not
include all relevant slow hydrodynamic modes. As may be inferred from a
paper by Belitz \cite{belitz1984}, however, projective results may at least be viewed
as reliable lower bounds to the transport coefficients. Furthermore, in
our projection we do not only keep, say, the particle and the energy
current but all individual occupation numbers of momentum modes.

\section{Introduction of the model} \label{sec-model}

In this paper we investigate 1-d models of weakly mutually interacting spinless fermions. I.e., we consider fermions on a periodic 1-d crystal lattice. The total system consists of a hopping model, which describes the non-interacting fermions, and a nearest neighbor interaction term. The Hamiltonian reads

\begin{eqnarray}
H=J\sum_{n=1}^{N} \Big[ \frac{1}{2}(a^{\dagger}_{n}a_{n+1}+ b\, a^{\dagger}_{n}a_{n+2} + \mathrm{h.c.} )\nonumber\\
+ \Delta \, a^{\dagger}_{n} a_{n}a^{\dagger}_{n+1} a_{n+1} \Big] \ .
\label{hampos}
\end{eqnarray}

In the hopping term we incorporate nearest neighbor hopping determined by the parameter $J$, according to a standard tight binding model, and also allow for some next-nearest neighbor hopping measured by the parameter $b$. $\Delta$ corresponds to the interaction strength. $N$ is the number of lattice sites. One may diagonalize the non-interacting system via Fourier transformation and arrives at

\begin{eqnarray}
H= \sum_{k} \varepsilon_{k}a^{\dagger}_{k}a_{k} 
+ \frac{1}{2}\sum_{k,l,q} \frac{W(q)}{N} \, a^{\dagger}_{k+q} a^{\dagger}_{l-q}a_{k} a_{l}  \ , \nonumber \\
\varepsilon_{k}=J(\mathrm{cos}(k)+b\, \mathrm{cos}(2k))\ , \quad W(q)=-2J\Delta \, \mathrm{e}^{-\imath q}.
\label{ham}
\end{eqnarray}

This Hamiltonian describes an interacting quantum gas model with the dispersion relation of the non-interacting fermions $\varepsilon_{k}$ and the interaction parameter $W(q)$. (The interaction of course obeys quasi momentum conservation with the quasi momentum transfer $q$.) For vanishing next-nearest neighbor hopping ($b=0$), $\varepsilon_{k}$ corresponds to a cosine band, as usually obtained from a tight binding model.

This model may describe, e.g., weakly interacting electrons on some 1-d atomic wire, possibly deposited on some substrate, or, e.g., as further discussed below, interacting spin systems, like for example an anisotropic spin 1/2 Heisenberg chain (by implementing some mapping of spins onto spinless fermions via Jordan-Wigner-transformation).

\section{Transport and Diffusion Coefficient} \label{sec-trans}

In this section we investigate the transport behavior of the above introduced system by directly following a scheme described in \cite{bartsch2010-1}. Basically, this perturbative approach is composed of two steps. Firstly, we map the underlying quantum dynamics of a certain set of "occupation number deviations from equilibrium" in momentum space onto some pertinent master equation, which may in some sense by interpreted as a scattering term of a corresponding linear(ized) Boltzmann equation. Further information about the mapping of quantum dynamics onto Boltzmann equations may be found in \cite{bartsch2010-1} and references therein. In the approach at hand the number of variables of the master equation is determined by the (finite) number of lattice sites $N$. The occupation number deviations from thermal equilibrium are described by operators
\begin{eqnarray}
\Delta_{{j}}:=(1-f_{{j}})a_{j}^{\dagger}a_{j}-f_{{j}}a_{j}a_{j}^{\dagger}=
a_{j}^{\dagger}a_{j}-f_{{j}}\ , 
\end{eqnarray}
where $f_{{j}}$ actually corresponds to the equilibrium Fermi distribution.

Note that throughout this paper we focus on the case of infinite temperature, i.e., in the following we assume $f_{{j}}= 1/2$.

As dynamical variables we consider the time dependent expectation values of these operators, i.e.,
$d_{{j}}(t):=$Tr$\{\Delta_{{j}}\rho(t)\}$,
where $\rho(t)$ is the density operator which describes the actual state of the system as resulting from the unitary Schr\"odinger dynamics. 

Our approach is based on a pertinent projection operator technique (for details see \cite{breuer2007}). Within this framework one has to define a suitable projection operator, which maps the current system's state $\rho(t)$ onto a density matrix-like object that only contains the variables of interest (here $d_{{j}}(t)$) as time dependent quantities. In this context the projection is chosen as
\begin{eqnarray}
\mathcal{P}\rho(t)=\rho^{\mathrm{eq}}+\sum_{{j}}\frac{\rho^{eq}\Delta_{{ j}}}{\text{Tr}\{\rho^{eq}\Delta^{2}_{{ 
j}}\}} d_{j}(t) \ .
\label{proj}
\end{eqnarray}

The resulting master equation reads 

\begin{equation}
\dot{d}_{i}(t)= \sum_{k\neq i} R_{i k}(t) d_{k}(t)-\sum_{k \neq i} R_{k i}(t) d_{i}(t) \ ,
\label{mastereq}
\end{equation}
where the corresponding rates ($T=\infty$, $k\neq i$) are given by 

\begin{eqnarray}
&&\hspace{-0.5cm}R_{{ki}}(t)=\int_{0}^{t}\text{d}\tau \frac{2}{\hbar^{2}}\frac{1}{4}\sum_{l} \nonumber\\
&&\hspace{-0.5cm}(\text{Re}(W({i}-{k}))-\text{Re}(W({k}-{l})))^{2}\, \frac{1}{N ^{2}}\nonumber\\
&&\hspace{-0.5cm}\cos [\frac{1}{\hbar}(\varepsilon_{{i}}+\varepsilon_{{l}}-\varepsilon_{{k}}-\varepsilon_{{i-k+l}})\tau] \nonumber\\
&&\hspace{-0.5cm}-\frac{1}{2}(\text{Re}(W({l}-{i}))-\text{Re}(W({k}-{l})))^{2}\, \frac{1}{N ^{2}}\nonumber\\
&&\hspace{-0.5cm}\cos [\frac{1}{\hbar}(\varepsilon_{{k}}+\varepsilon_{{i}}-\varepsilon_{{i+k-l}}-\varepsilon_{{l}})\tau]  \ .
\label{el-el}
\end{eqnarray}

I.e., every element of the rate matrix (\ref{el-el}) is given by an integral over a certain correlation function. We especially intend to evaluate (\ref{el-el}) in the limit of small interactions and large time $t$.

Our approach to transport applies if all rates become constant in this limit for times larger than some correlation times $\tau_{c,ik}$.
If this is the case, we may view the rate matrix $R_{k i}$ as the collision term of a Boltzmann equation from which a diffusion coefficient may be obtained using a type of Chapman-Enskog-like approach, see \cite{bartsch2010-1}. We omit any details here for brevity.
We find for the diffusion coefficient (to the order of $\Delta^{-2}$ and for $T=\infty$)
\begin{equation}
D =-\frac{1}{N}v_{k}R^{-1}_{k i}v_{i} \ ,
\label{difco}
\end{equation}
where $R^{-1}$ is the inverse matrix to $R$ neglecting the eigenspace corresponding to the equilibrium distribution $f_{j}$.
$v_{i}$ is the momentum dependent velocity/current vector, the individual velocities are given by the slope of the dispersion relation of the non-interacting system $v_{i}=\partial \varepsilon_{i}/\partial i$. (Hereby we follow the common idea of identifying particle velocities with group velocities.)

Although we consider finite lattices, the obtained diffusion coefficients have to be robust against upscaling of the system in terms of $N \rightarrow \infty$ in order that our results are reasonable for large ("realistic") crystals.

\section{Numerical Results for the Diffusion Coefficient for ${\bf b\neq 0}$} \label{sec-numerics}

For $b\neq 0$ the analytical evaluation of the long time limit of the rates (\ref{el-el}) is not directly accessible. Instead we determine $R_{ki}(t)$ numerically for finite time $t$. 
(Numerics, which we omitted here, suggest that the rate matrix elements $R_{ki}(t)$ may be expected to become time independent after some correlation times $\tau_{c,ik}$.)
Based on this time dependent rate matrix we may determine an initially also time dependent diffusion coefficient $D(t)$ by directly evaluating formula (\ref{difco}), which is our first main numerical result. 

%%%%%%%%%%%%%%%%%%%%%%%%%% FIG. 2 %%%%%%%%%%%%%%%%%%%%%%%%%%%%%%%%%
\begin{figure}[htb]
\centering
\includegraphics[width=6.4cm]{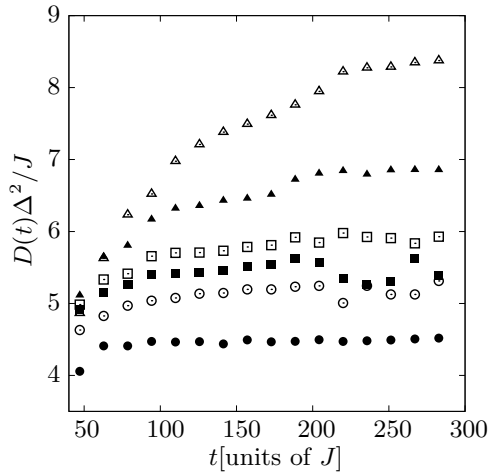}
\hspace{1.0cm}
\caption{Time dependent diffusion coefficients for system with next-nearest neighbor hopping $b=0.2$ (white triangles), $b=0.3$ (black triangles), $b=0.4$ (white squares), $b=0.5$ (black squares), $b=0.6$ (white circles), $b=0.8$ (black circles). $N=500$, $\Delta=0.01$.} 
\label{dift}
\end{figure}
%%%%%%%%%%%%%%%%%%%%%%%%%%%%%%%%%%%%%%%%%%%%%%%%%%%%%%%%%%%%%%%%%%%%

Fig. \ref{dift} demonstrates these time dependent diffusion coefficients for different next-nearest neighbor hoppings $b$. One finds that for all analyzed $b$ the diffusion coefficients $D(t)$ reach plateau levels and therefore become approximately time independent for large enough times. The constancy of $D$ of course indicates that the elements of the rate matrix $R_{ki}$ have converged, too. I.e., within this framework we are able to numerically determine finite long time diffusion coefficients for non-vanishing next-nearest neighbor hopping $b \neq 0$.

We essentially find the following two quantitative results. The long time diffusion coefficients become larger for smaller $b$. This finding clearly demonstrates the quantum nature of the model and the approach, because it seems counter-intuitive in the sense that it does not fit the (classical) perspective that an increased hopping would lead to a higher diffusion.

%%%%%%%%%%%%%%%%%%%%%%%%%% FIG. 2 %%%%%%%%%%%%%%%%%%%%%%%%%%%%%%%%%
\begin{figure}[htb]
\centering
\includegraphics[width=6.4cm]{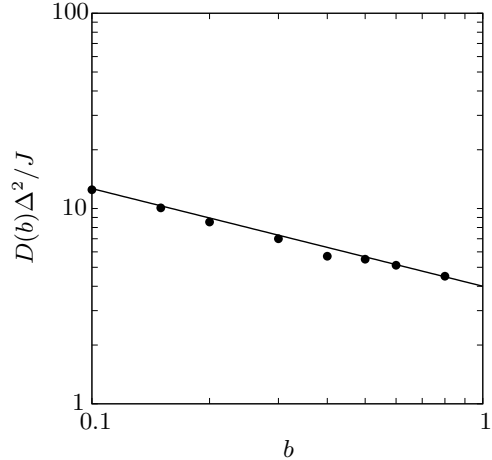}
\hspace{1.0cm}
\caption{Long time diffusion coefficients versus next-nearest neighbor hopping $b$. The numerics (points) are well approximated by the function $4/\sqrt{b}$ (solid line). $D(b)$ diverges for $b=0$.} 
\label{difb}
\end{figure}
%%%%%%%%%%%%%%%%%%%%%%%%%%%%%%%%%%%%%%%%%%%%%%%%%%%%%%%%%%%%%%%%%%%%

Additionally, Fig. \ref{difb} shows the dependence of the long time diffusion coefficients on the next-nearest neighbor hopping $b$. The plot suggests a dependence $D(b)\propto b^{-1/2}$ (, although we do not have an analytical proof for this type of relation), which would mean that the diffusion coefficient diverges for $b=0$.

Secondly, the time, after which the diffusion coefficient is approximately constant, becomes larger for smaller $b$. I.e., to analyze the interesting limit $b \rightarrow 0$ (see below) we have to evaluate longer and longer times to obtain the plateau level (, which goes along with larger chain lengths $N$; recall that the results should be robust against upscaling of $N$). 

Since for $b=0$ the plateau is reached only at infinite times and the plateau height is expected to be infinite, which is in some sense equivalent, we cannot numerically extract finite diffusion coefficients in this case. The findings may rather suggest that the transport is possibly non-diffusive for $b=0$. Some analytical treatment of this case can be found below in Sec. \ref{sec-analytics}.

At this point this result should be compared to a recent, related result
by Steinigeweg et al. \cite{steinigeweg2010}. They consider the same model
class but only for $b=0$ and employ a related projection technique.
However, other than in the work at hand, they do not keep all individual
momentum occupation numbers but project onto a single variable, namely
the current $j=\sum_i v_i a_i^+a_i$. Doing so they find a finite
diffusion coefficient emerging at short time scales for all $\Delta$ including arbitrarily small ones.
We checked that this scenario gradually transforms into the picture we find as more and
more observables in addition to the current are kept in the projection,
i.e., time scales become longer and diffusion constants increase. This,
however, is, according to Belitz \cite{belitz1984}, not a contradiction since
results from projections onto subsets of observables yield reliable
lower bounds to the true transport coefficients. Thus we conclude, in
spite of the findings by Steinigeweg et al. that there is no regular
diffusive transport for $b=0$.

Concerning the short time/strong interaction behavior of the diffusion coefficients we find that the graphs for different $b$ in Fig. \ref{dift} intersect at some time $t$. It may be thinkable that the behavior of the diffusion coefficients reverses for small times in the way that the diffusion coefficient for small times becomes larger for larger $b$, which would be in agreement with (or at least not contradictory to) the above mentioned intuitive expectation. However, since in our approach the rate matrix is not converged out at those times, we cannot assure that the short time values are quantitatively correct, they rather serve as an indicator for the overall qualitative behavior. For this regime also see \cite{steinigeweg2011}.

\section{Analytical Considerations on Transport for ${\bf b=0}$} \label{sec-analytics}

The special case of vanishing next-nearest neighbor hopping $b=0$ is of particular interest because the regarded model of spinless fermions (\ref{ham}) is effectively equivalent to an anisotropic spin-1/2 Heisenberg chain. The corresponding spin Hamiltonian

\begin{equation}
H = J\sum_{n=1}^{N} \Big[ \frac{1}{2}[ s^{+}_{n}s^{-}_{n+1} \\
+ \, \mathrm{h.c.} ]
+ \Delta \, s^{z}_{n} s^{z}_{n+1}\Big]
\end{equation}

may be transformed into the Hamiltonian (\ref{hampos}) with $b=0$ via a suitable Jordan-Wigner-transformation. Note that we consider here the case of small $\Delta$, which rather corresponds to the case of strong anisotropy in the spin picture.

Recall that in our numerical evaluation in Sec. \ref{sec-numerics} we have found diverging long time diffusion coefficients for $b=0$ which we suggested to interpret as an indicator for possibly non-diffusive transport. If this is actually the case, one would expect to find a number of conserved quantities, at least some of which should have a significant overlap with the current. (Since the model is integrable if (and only if) $b=0$ (see Sec. \ref{sec-introduction}), there are a large number of conserved quantities for $b=0$. The transition from non-integrability to integrability may here be related to the transition from diffusive to non-diffusive transport.) 

For $b=0$ one may analytically determine the rates $R_{ki}$ in the limit $t\rightarrow \infty$. 
Assuming that the momentum modes are lying densely in $k$-space (i.e., assuming long chains,) one may substitute the sum over $l$ in (\ref{el-el}) by a corresponding integral, i.e.,
\begin{equation}
\sum_{l} \rightarrow \frac{N}{2\pi}\int_{l}\ ,
\end{equation}
and regard the energy terms 
$e_{+}(l):=(\varepsilon_{{i}}+\varepsilon_{{l}}-\varepsilon_{{k}}-\varepsilon_{{i-k+l}}))$
in the first and
$e_{-}(l):=(\varepsilon_{{k}}+\varepsilon_{{i}}-\varepsilon_{{i+k-l}}-\varepsilon_{{l}})$ 
in the second part of the sum as continuous functions of $l$. 
Routinely exploiting the properties of the $\text{sinc}$-function one may carry out the time integration in (\ref{el-el}) and finds that in the limit $t\rightarrow \infty$ only the exactly energy conserving scattering processes contribute, i.e., only addends featuring ($e_{+}= 0$) in the first and ($e_{-}= 0$) in the second term. 

\begin{figure}[htb]
\centering
\includegraphics[width=7.0cm]{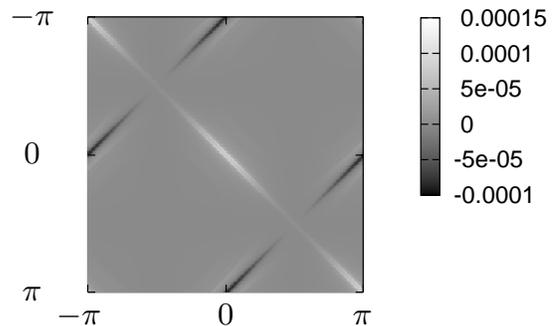}
\caption{Finite time sketch of the rate matrix $R_{j k}$ for $b=0$ without diagonal elements, the labels denote the corresponding momentum mode numbers. There are large (absolute) negative scattering matrix elements (here black) between modes near lines with $j+k=\pi$ and $j+k=-\pi$, which originate from the ``minus'' term in \ref{el-el}. Parameters: $N=200$, $J=-1$, $\Delta = 0.01$, $T=\infty$.} 
\label{rmat}
\end{figure}

The second ``minus'' term yields diverging $\delta$-like contributions on lines, with respect to a graphical illustration of the scattering operator, defined by $j+k=\pi$ and $j+k=-\pi$, but only exactly on those lines. For all other $j,k$, even in the direct vicinity of those lines, the ``minus'' term vanishes, i.e., the corresponding contribution to the rate matrix from the second term may be written as 
$R^{-}(j,k)\propto -\delta(j+k-\pi)- \delta(j+k-\pi) +\delta(j-k)$ for $t\rightarrow\infty$.
This feature may be illustrated by a finite time sketch of the total rate matrix $R_{j k}$ (without diagonal elements) (see Fig. \ref{rmat}). One finds large negative matrix elements in the vicinity of the above mentioned lines, for larger $t$ the negative contributions move closer to the lines, which eventually leads to the $\delta$-like structure.

%%%%%%%%%%%%%%%%%%%%%%%%%% FIG. 2 %%%%%%%%%%%%%%%%%%%%%%%%%%%%%%%%%
%%%%%%%%%%%%%%%%%%%%%%%%%%%%%%%%%%%%%%%%%%%%%%%%%%%%%%%%%%%%%%%%%%%%

Although these terms are problematic for the determination of the complete rate matrix, they yield no contribution to the diffusion coefficient (\ref{difco}) because of symmetry arguments (see below) and may therefore be omitted in the transport investigations addressed in this paper. 

The complete rate matrix for $b=0$ features mirror symmetry to the band middles, i.e., lines with $k=\pi/2$ and $k=-\pi/2$, because the $\text{cos}$-function simply passes into its negative under inversion on those lines. That is, every eigenvector of the rate matrix has to transform into itself or its negative under mirror imaging on those lines and the eigenvectors may accordingly be classified by, say $+$ and $-$. The matrix therefore separates into these symmetry subspaces, i.e., if one chooses a matrix representation where all basis vectors feature either $+$ or $-$ symmetry, there is no coupling between the two subspaces, the corresponding matrix elements have to vanish.
Note that the current features $+$ symmetry.
For any $+$ vector $d(j)$ one finds that $d(j)=d(\pi -j)$ for $j>0$ and $d(j)=d(-\pi -j)$ for $j<0$.
That is, the multiplication of the ``minus'' part $R^{-}(j,k)$ with a $+$ vector 
\begin{eqnarray}
\tilde{d}(j)&=&\int dk \, R^{-}(j,k) d(k) \nonumber\\
&=& - d(-j-\pi)-d(-j+\pi)+d(j)
\end{eqnarray}
is equal to $0$, since two terms cancel each other, depending on whether $j$ is positive or negative, and the remaining term does not contribute to the integration because the corresponding $k$-argument is not in the first Brillouin zone.
Therefore $R^{-}(j,k)$ gives no contribution to the sum evaluated in (\ref{difco}), which means that $R^{-}(j,k)$ can be neglected.

From the remaining first ``plus'' term in (\ref{el-el}) one finally obtains from collecting and evaluating all the zeros of the respective argument in (\ref{el-el})
\begin{equation}
R^{+}(j,k)\propto \frac{1}{N}\frac{(\text{cos}(k-j)+\text{cos}(k+j))^{2}}{\vert \text{sin}(k)- \text{sin}(j) \vert}
\label{rateinf}
\end{equation}
for $t=\infty$. One recognizes that the rate matrix $R^{+}(j,k)$ is here a markedly complicated object, mainly because of two reasons.

There are diverging elements of $R^{+}(j,k)$ in the vicinity of lines defined by $j+k=\pi$ and $j+k=-\pi$. For larger and larger systems ($N\rightarrow\infty$), when the momentum modes become denser and denser in $k$-space, there are more and more elements in the direct vicinity of those lines, which therefore become arbitrarily large. From this one may expect that at least some eigenvalues of $R^{+}(j,k)$ diverge in the limit $N\rightarrow\infty$.

\begin{figure}[htb]
\centering
\includegraphics[width=6.4cm]{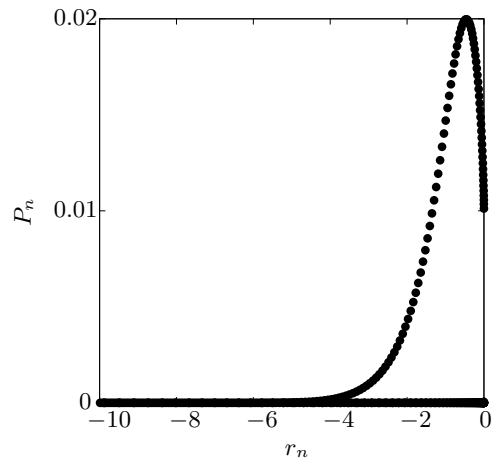}
\hspace{1.0cm}
\caption{Overlap $P_{n}:= (\vec{j}\cdot \vec{r}_{n})^{2}$ of the current $\vec{j}$ with the eigenvectors $\vec{r}_{n}$ of the matrix $R^{+}(j,k)$ versus corresponding ``rescaled'' eigenvalue ($r_{n}\Delta^{2}$ are eigenvalues of $R^{+}(j,k)$) for time $t\rightarrow \infty$. There is a finite overlap with (approximately) conserved quantities for $r_{n}\approx 0$, i.e., parts of the current decay arbitrarily slowly. Parameters: $N=800$, $J=-1$, $T=\infty$.} 
\label{ewpl}
\end{figure}

Secondly, there are whole lines in the matrix (\ref{rateinf}) defined by $j,k = -\pi /2$ and $j,k = \pi /2$  where all matrix elements are $0$, indicating conserved quantities.
I.e., $d(j)\propto \delta (j- \pi /2)+ \delta (j+ \pi /2)$ is a conserved quantity, which features a non-singular but finite overlap with the current. This causes an infinitely small portion of the current to decay infinitely slowly (see Fig.\ref{ewpl}.) In \cite{sirker2009} it is pointed out that a non-diffusive behavior of the system would require a non-local conservation law. Note, however, that the above conserved quantity features such a spatially non-local structure.

To take a closer look at the concrete transport behavior we evaluate the overlap of the current/velocity vector with the eigenvectors of $R^{+}(j,k)$ in dependence of the corresponding eigenvalues.

%%%%%%%%%%%%%%%%%%%%%%%%%% FIG. 2 %%%%%%%%%%%%%%%%%%%%%%%%%%%%%%%%%
%%%%%%%%%%%%%%%%%%%%%%%%%%%%%%%%%%%%%%%%%%%%%%%%%%%%%%%%%%%%%%%%%%%%

Firstly, Fig.\ref{ewpl} shows that the current does not decay monoexponentially, although an average relaxation rate may be identified. In particular there is a finite overlap between the current and conserved (arbitrarily slowly relaxing) quantities, i.e., an infinitely small portion of the current does not decay at all. This can explain why we are not able to find converging diffusion coefficients according to formula (\ref{difco}) and why the numerically calculated curve $D(t)$ (cf. Fig. \ref{dift}) does not exhibit a plateau for $b=0$.  Fig.\ref{ewpl} also demonstrates that there are many quantities which have no overlap with the current.

Within the scope of the master equation dynamics (\ref{mastereq}) the current autocorrelation function $C(t):=\langle j(t)j \rangle$ decays multiexponentially, i.e.,
\begin{equation}
C(t)=\sum_{n} P_{n} \mathrm{e}^{r_{n} \Delta^{2} t} \ ,
\label{csum}
\end{equation}
where $r_{n}\Delta^{2}$ are eigenvalues of the rate matrix $R^{+}(j,k)$ ($r_{n}\leq 0$) and $P_{n}$ corresponds to the overlap of the current vector with the corresponding eigenvector $\vec{r}_{n}$ of $R^{+}(j,k)$ defined by $P_{n}:= (\vec{j}\cdot \vec{r}_{n})^{2}$. For large systems one may substitute the sum by an integral
\begin{equation}
C(t)=\int P(r) e^{r \Delta^{2} t} dr 
\label{cint}
\end{equation}
with continuous eigenvalues $r\Delta^{2}$ and the function $P(r)$ essentially given by Fig. \ref{ewpl}. The long time behavior is mainly determined by the regime of small $r$. Therefore we approximate $P(r)$ by the corresponding Taylor expansion around $r\approx 0$
\begin{equation}
P(r)\approx P(0) + \sum_{n=1}^{\infty} c_{n} r^{n} 
\label{ptaylor}
\end{equation}
with some coefficients $c_{n}$. Inserting (\ref{ptaylor}) into (\ref{cint}) one finds that for times $t$ larger than $1/\Delta^{2}$ the behavior of $C(t)$ is dominated by the term $C(t)\approx P(0)/ (\Delta^{2} t)$ which results from the zeroth order term of (\ref{ptaylor}). All other contributions decay with $t^{-2}$ or higher orders in $t$ or exponentially with $t$.
Although $C(t)$ is essentially a sum of exponential functions, it does not behave exponentially in this limit.
A time dependent diffusion coefficient $D(t)$, which may be decomposed as
\begin{equation}
D(t)=\int_{0}^{1/\Delta^{2}} C(t') dt' + \int_{1/\Delta^{2}}^{t} C(t') dt' \ ,
\label{difint}
\end{equation}
is dominated by the second term for times larger than $1/\Delta^{2}$. In this limit $D(t)$ behaves logarithmic as
\begin{equation}
D(t)\propto (1/ \Delta^{2} ) \text{ln}(\Delta^{2} t) \ , 
\label{diflog}
\end{equation}
and consequently does not become constant for $t\rightarrow \infty$.

Also note that the integral over the current-autocorrelation function as, e.g., given in (\ref{difint}) is directly related to the time dependent growth of the mean square displacement (MSD)
of some initial distribution, the corresponding connection has been established in \cite{steinigeweg2009}.

Within the scope of our approach the result (\ref{diflog}) would lead to the conclusion that the transport is not regularly diffusive and neither really ballistic (in this case one would find $D(t)\propto t$), but rather something in between, which may, e.g., be termed ``anomalous diffusion'' (cf. also \cite{dahlqvist1996}). This characteristic behavior arises from the fact that the function $P(r)$ (cf. Fig. \ref{ewpl}) exhibits a finite value at $r=0$. Ballistic transport would require a peak at $r=0$. 
From (\ref{diflog}) one may suspect that the MSD would grow proportional to $t \, \text{ln}(t)$. 

However, our approach is restricted to diffusion coefficients of the order of $\Delta^2$, i.e., there could possibly be diffusive behavior scaling with $\Delta^{4}$ or higher orders.

\section{Conclusion} \label{sec-conclusion}

We demonstrated that the quantum dynamics of momentum mode occupation numbers in a 1-d quantum model for weakly interacting fermions may be mapped onto a master equation, the corresponding rate matrix can serve as collision term of a linear(ized) Boltzmann equation. We numerically found that adequately calculated diffusion coefficients become constant in the limit of large times and large systems for finite next-nearest neighbor hopping. These long time diffusion coefficients are larger for smaller next-nearest neighbor hopping. For vanishing next-nearest neighbor hopping the diffusion coefficients diverge. In this case we have analytically shown that arbitrarily small parts of the current do not relax because there is a finite overlap between the current and (approximately) conserved quantities. This result suggests anomalous diffusive transport behavior for the model featuring no next-nearest neighbor hopping.

\begin{acknowledgments}
We sincerely thank R. Steinigeweg and H. Niemeyer for fruitful discussions.
\end{acknowledgments}

\end{document}